\documentclass[%
 aip,
 jcp,%
 amsmath,amssymb,
 preprint,%
 floatfix
]{revtex4-1}

\usepackage{graphicx}
\usepackage{dcolumn}
\usepackage{bm}
\usepackage{epstopdf}
\usepackage[version=3]{mhchem} 
\usepackage{threeparttable}

\begin{document}

\title[Cellulose-NaCl interactions]{Sodium Chloride interaction with
solvated and crystalline cellulose :\\ sodium ion affects the tetramer and
fibril in aqueous solution}

\author{Giovanni Bellesia}
\affiliation{Theoretical Biology and Biophysics Group, Center for Nonlinear Studies, Los Alamos National Laboratory, Los Alamos, NM 87545, USA}
\email{giovanni.bellesia@gmail.com}
\author{S. Gnanakaran}
\affiliation{Theoretical Biology and Biophysics Group, Los Alamos National Laboratory, Los Alamos, NM 87545, USA}

\date{\today}

\begin{abstract}
Inorganic salts are a natural component of biomass which have a significant
effect on the product yields from a variety of biomass conversion processes.
Understanding their effect on biomass at the microscopic level can help
discover their mechanistic role. We present a study of the effect of aqueous
sodium chloride (NaCl) on the largest component of biomass, cellulose, focused
on the thermodynamic and structural effect of a sodium ion on the cellulose tetramer, and fibril. 
Replica exchange molecular dynamics simulations of a cellulose tetramer reveal a number of preferred 
cellulose-Na contacts and bridging positions. Large scale MD
simulations on a model cellulose fibril find that \ce{Na+} perturbs the
hydroxymethyl rotational state population and consequently disrupts the `native'
hydrogen bonding network.
\end{abstract}

\maketitle

\section{\label{sec:intro}Introduction}

Inorganic salts are an integral part of any biomass material.
Indeed, plant biomass naturally contains alkali and alkaline earth metals, including
potassium, sodium, phosphorus, calcium, and
magnesium.\cite{Mohan2006,Patwardhan2009,Patwardhan2011} While the total
mineral content is less than 1 \%, the effect, either adverse or favorable, on
biomass degradation and product distribution is significant. Removal of these
cations would add cost\cite{Scott2001} to biofuel production, which is already
more costly than conventional petroleum-based fuels. Thus, it is preferable to 
understand how the inorganic material interacts with the biomass to better steer 
conversion toward more valuable products. In particular, a detailed
understanding of the interactions between inorganic salts and biomass is of
central importance for devising efficient biomass degradation protocols for
bioenergy production.

In fast pyrolysis, the presence of inorganic salts has been shown to increase
the yield of lower-value low-molecular weight species and adversely affects the
formation of levoglucosan (the main pyrolysis product of pure cellulose).
\cite{Patwardhan2010} Conversely, in steam gasification of biomass-derived
charcoal, Li, Na and K chlorides have been used as catalysts to increase gas
yield and to reduce the operational temperature.\cite{Encinar2001} The
beneficial effect of inorganic salts has also been outlined in recent studies
showing that the presence of NaCl increases the yield of levulinic acid in
acid-catalyzed depolymerization of cellulose in
water.\cite{vomStein2010,Potvin2011} 
Those studies are especially relevant as
they combine low-price NaCl with relatively mild reaction conditions and low
temperatures ($T=373-473$ K). In those acid catalysis experiments the NaCl
concentration varies in the range of $5-50 \%$ wt. The peak in the sugars'
yield has been obtained with NaCl concentrations in the range of $20-30 \%$ wt.
It is speculated that NaCl contributes to the destabilization of the
highly-structured hydrogen bond network in cellulose, thereby increasing the
efficiency of the cellulose hydrolysis process.

In our study we employ classical computational methods to study the
atomistic details of the interactions of NaCl oligomeric and crystalline fibrillar cellulose. 
Oligomeric and crystalline fibrillar cellulose were studied via replica exchange molecular dynamics 
and via single temperature molecular dynamics (MD) simulations respectively. 

Our results show that there are multiple positions for \ce{Na+} to associate with
hydroxyl groups of a pyranose ring within thermal energy ($k_BT$) at
temperatures relevant to biomass conversion. Thus, it is expected that these
ions could affect many different reactions and thus have wide-ranging effects
on decomposition of biomass.

\section{\label{sec:meth}Computational Methods}

We performed replica exchange Langevin dynamics (LD) and single temperature
NPT--LD simulations to analyze the equilibrium dynamics of a cellulose tetramer
and of a model fibril of crystalline cellulose in NaCl aqueous solution,
respectively. Replica Exchange LD simulations\cite{Swensen1986,Sugita1999} were
carried out in a simulation box of dimensions $L_x=38.8$  \AA\, $L_y=43.1$ \AA\,
and $L_z=34.4$  \AA\ containing $1839$ water molecules and $5$ NaCl pairs. The
cellulose-NaCl ratio 
in the cellulose tetramer system 
corresponds to a $40$ wt\% 
concentration, which is within the range of the experimental wt\% concentrations 
used in salt-assisted acidcatalysis experiments of cellulose decomposition.\cite{vomStein2010,Potvin2011}
Forty-eight replicas were used corresponding to $48$ temperatures in the
interval $[298-550]$ K. The total simulation time per replica was $100$ ns with
\textit{replica swaps} performed every $40$ ps and a \textit{swap acceptance
ratio} varying between $0.1$ and $0.3$. The last $50$ ns were considered for
data analysis. The cellulose tetramer was capped with reducing and non--reducing
ends.\cite{glycam-url,Shen2009} The tetramer and the ion pairs were spatially
constrained within a sphere of radius $r_s=15$ \AA\, centered in the water box
center.\cite{GioBJ2009} In more detail, spherical harmonic boundary conditions
were enforced on the tetramer and the ion pairs by means of a single potential
function,
\begin{equation}
E_{s}=\left\{ \begin{array}{rl} 
k_{s}(|\mathbf{r}_i-\mathbf{r}_c|-\mathbf{r}_s)^2 &\mbox{if
$|\mathbf{r}_i-\mathbf{r}_c|>r_s$}, \\ 0 &\mbox{otherwise}
\end{array} \right.
\end{equation}
where $k_s=10$ kcal/mol, $\mathbf{r}_i$ is the current position of atom i and
$\mathbf{r}_c$ is the center of the sphere. The potential $E_{s}$ was used to
increase the NaCl concentration ($\approx0.6$ M) without increasing the size of
the system and, therefore, the computational cost of the simulations.
Additional Replica Exchange LD simulations were carried out for a cellulose
tetramer in aqueous solution without NaCl under the same simulation conditions
used in the NaCl simulations. For these calculations, the total simulation time
per replica was $40$ ns and the last $20$ ns were considered for data analysis.
Two single temperature NPT--LD simulations on a crystalline I$_{\beta}$
cellulose fibril composed of $30$ octameric glucan chains were performed at
$T=298$ K and $T=400$ K. For the initial fibril conformation we considered an
equilibrated structure from previous LD simulations.\cite{ShiGio2011} Each
octameric cellulose chain was covalently connected to its periodic image along
its main axis in order to mimic an infinitely long cellulose fibril.\cite{ShiGio2011} 
The fibril was solvated in a rectangular box containing $6660$ water molecules and $15$
NaCl pairs. The cellulose-NaCl ratio 
in the cellulose fibril system
corresponds to $2$ wt\%, which is within
the concentration range present in real biomass feedstock ($0.5-5\%$) and
relevant to fast pyrolysis experiments.\cite{Patwardhan2010} The $2$ wt\%
concentration used in our simulations is close to the lowest concentration used
in acid cataysis experiments ($5\%$ wt).\cite{vomStein2010,Potvin2011} The use
of this relatively low NaCl concentration is justified by the fact that for
higher concentrations (both in terms of weight percent and molarity), ions tend
to form small unphysical clusters that strongly limit the NaCl reactivity and
negatively affect the statistical sampling of the NaCl-cellulose interactions.
This is a well known problem that arises when classical force fields are
used.\cite{Amber99,Pappu2007}

The solvated cellulose fibril underwent first a local optimization, followed by
a short ($0.5$ ns) NPT--MD simulation, where the temperature was gradually
increased from $100$ K to $298$ K. This initial optimization-thermalization
simulation was followed by two separate $100$ ns long NPT--MD simulations at
$T=298$ K and $T=400$ K, respectively. The pressure P was $1.01325$ bar. The
first $10$ ns of the $100$ ns run were considered as the initial equilibration
time and the remaining $90$ ns as the production time. During the $100$ ns
simulation runs we applied a cylindrical boundary potential (oriented along the
cellulose fibril main axis) analogous to the one shown in Equation 1 using a
radius $r_{cyl}=35$ \AA\ and an infinite length. We found that the magnitude of 
both the spherical and the cylindrical potential is negligible when compared to
the total potential energy of our systems ($<0.01\%$ of the total potential
energy). Hence, we do not expect the presence of the boundary potentials to 
affect our results and conclusions. In addition, we used the same protocol to
run a $50$ ns simulation on a fibril in pure water at $T=400$ K. In all
simulations, the time step was fixed at $2.0$ fs. The covalent bonds involving
hydrogen atoms were constrained by means of the SHAKE
algorithm.\cite{Ryckaert1977} We used the NAMD software
package\cite{Phillips2005} with the GLYCAMO6\cite{Kirschner2008} force field
and the TIP3P explicit water model.\cite{tip3p} 
The reliability of the GLYCAM force field for solvated and crystalline cellulose systems 
has been assessed in a number of recent publications.\cite{Shen2009,ShiGio2011,Gioamm2011,Matthews2012comparison}
A Langevin thermostat and
Nose--Hoover Langevin barostat with a stochastic component were used to control
the temperature and the pressure, respectively.\cite{Feller1995,Martyna1994}
The damping coefficient for the Langevin integrator was set to $1.0$ ps$^{-1}$, 
while for the Nose--Hoover Langevin barostat we applied an oscillation period of
$200$ fs and a damping period of $100$ fs. The cutoff for the non-bonded
interactions in the coordinate space was fixed at $10.0$ \AA. All the
simulations were performed under periodic boundary conditions, and the
long--range electrostatic interactions were calculated by using the Ewald
summation method with the particle mesh Ewald algorithm.\cite{Darden1993} The
particle mesh Ewald accuracy was fixed at $10^{-6}$, the order of the
interpolation functions on the grid was set to $4$ and the grid spacing was
$\approx1.0$ \AA. 

\section{\label{sec:res}Results and Discussion}

\subsection{\label{sec:res3}Cellotetraose Conformations}
\begin{figure}[!ht]
\centering
\includegraphics[angle=270,scale=0.5]{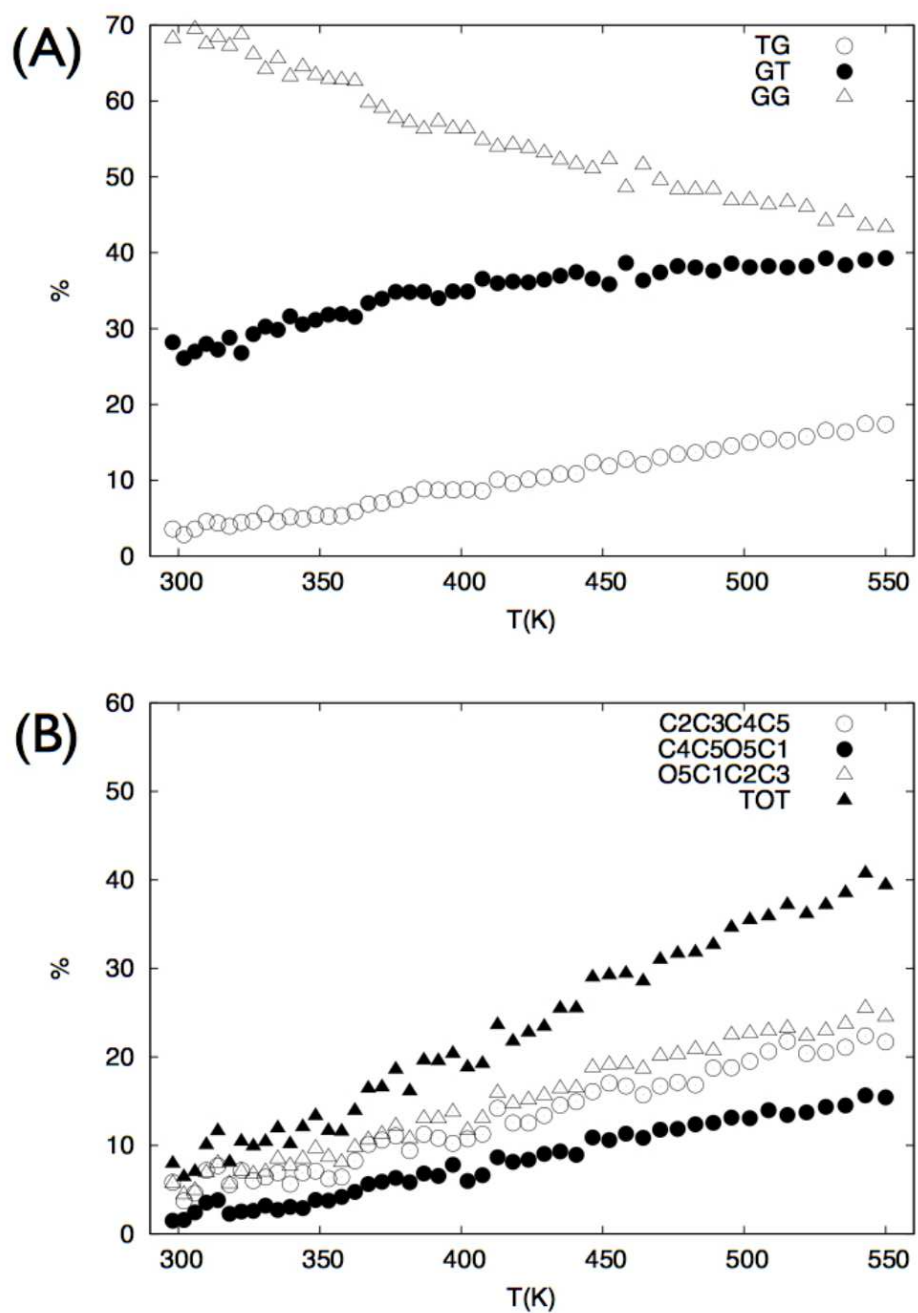}
\caption{(A) Percentage of the rotational state population for the
hydroxymethyl group as a function of temperature. (B) Percentage for the non
chair--like conformation of the glucose rings as a function of temperature.
Results are shown for each of the three dihedral angles defining the ring
conformation as well as for the total percentage of non-chair--like
conformations (black triangles).}
\label{fig:1}
\end{figure}

Conformations of a cellulose tetramer have been analyzed considering a set of
relevant dihedral degrees of freedom \cite{Shen2009}: the hydroxymethyl
rotational state \cite{Shen2009,ShiGio2011} and the three puckering angles
defining the conformation of the glucose ring ($C2C3C4C5$, $C4C5O5C1$,
$O5C1C2C3$). In Fig.~\ref{fig:1}(A) we plot the percentage for the three
hydroxymethyl rotational states ($tg$, $gt$ and $gg$) as a function of
temperature. At 298 K, the rotational population is mostly dominated
by the $gg$ state with the $gt$ state contributing as well. The statistical
weight of the $tg$ state (dominant in native crystalline cellulose) is quite
small at room temperature. Not surprisingly, the statistical weight of $tg$
increases with temperature as the thermal energy ``flattens'' the energy
barriers between the three hydroxymethyl states. In Fig. S1(A,B,C) in the supplementary 
information we show the residuals between $tg$, $gt$ and $gg$ data,
respectively, calculated as the difference between the percentages collected in
NaCl aqueous solution (Fig.~\ref{fig:1}(A)) and the analogous data collected
from the simulations of the cellulose tetramer in pure water. The normalized
sums of the residuals are $0.27\%$, $1.20\%$ and $-1.47\%$, for $tg$, $gt$ and
$gg$, respectively. These data show that when NaCl is present, the statistical
weight of the $gg$ state diminishes mostly at the expense of the $gt$ state
that becomes moderately more favorable than in pure water.

In Fig.~\ref{fig:1}(B) we plot the percentage of distorted, non-$^4C_1$
conformations (black triangles) defined by negative values of at least one of the dihedrals
$C2C3C4C5$, $C4C5O5C1$ and $O5C1C2C3$.\cite{Shen2009} We also show the partial
percentages for each of the three angles defining the ring conformation. The
percentage of distorted, non-chair--like conformations appears to increase
linearly with temperature. Comparison with simulations in pure water (Figure
S1(D)) shows that the statistical weight of the non-$^4C1$ conformations is
negligible when NaCl is present (normalized residual $=1.15\%$).

\subsection{\label{sec:res4}Cellotetraose--\ce{Na+} Interaction} 
To analyze the interaction of solvated cellulose with NaCl, and in particular
with \ce{Na+}, we first considered a set of radial distribution functions (RDFs)
related to ion--cellulose heavy atom interactions. The limited number of ions
in our system together with the lack of radial symmetry make the use of RDFs
problematic, especially when it comes to normalization at large distances.
Nevertheless, we make preliminary use of ion-cellulose RDFs to obtain
information about the location of the first minimum which defines the cutoff
distance for the ion--cellulose interactions (Fig. S2). A systematic
analysis of the RDFs reveals that the most relevant ion--cellulose interactions
involve \ce{Na+} and the hydroxyl oxygen atoms. The sharp peaks near $2.5$ \AA\
indicate that \ce{Na+} forms a well-defined coordination with hydroxyl groups. The
peak is stronger in the O3--\ce{Na+} case indicating that the coordination is
stronger in this case. With increasing temperature, the peaks lose intensity
and become slightly broader.

\begin{figure}[!ht]
\centering
\includegraphics[angle=0,scale=0.5]{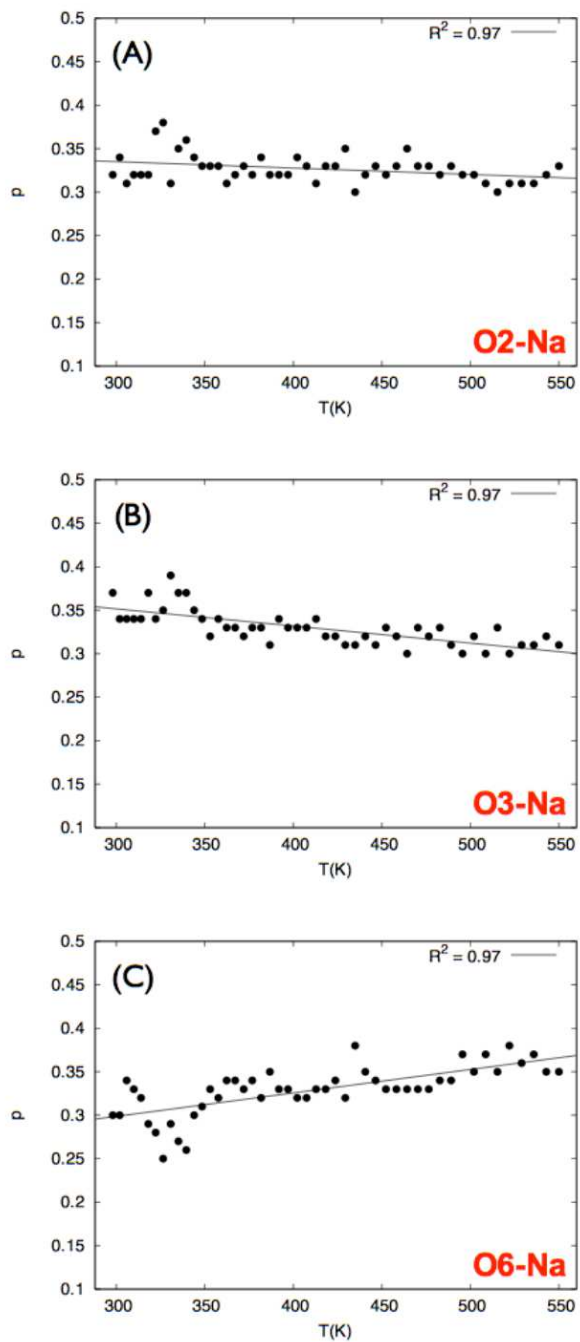}
\caption{Cellulose tetramer in NaCl aqueous solution.
Full temperature dependence of the relative probabilities of O2--\ce{Na+},
O3--\ce{Na+}, O6--\ce{Na+} contacts are shown in panels (A), (B) and (C),
respectively. O6 increases its contact probability to \ce{Na+} at higher
temperatures. At $\sim 410$ K, O6 contact probability to \ce{Na+} becomes the largest.}
\label{fig:2}
\end{figure}

The first minimum in those critical RDFs is typically well--defined and located
at a distance $r_c=3.2$ \AA. That distance represents an accurate cutoff
measure for the first coordination cell and for defining ``contacts'' between
\ce{Na+} and oxygen atoms in cellulose. In Table~\ref{tab:1}, we report the
relative probabilities for the ``contacts'' between \ce{Na+} ions and the relevant
oxygen atoms $O2$, $O3$ and $O6$ (hydroxymethyl oxygen) in cellulose at
$T=298$, $T=402$ and $T=502$ K. Oxygens that belong to the non--reducing end of
the cellulose tetramer are included in the calculations. Our results show that
$O3$ is slightly more prone to form contacts with \ce{Na+} than $O6$ and $O2$.
The full temperature dependence for the relative probabilities of the
$O2$--\ce{Na+}, $O3$--\ce{Na+} and $O6$--\ce{Na+} contacts is shown in
Fig.~\ref{fig:2}. The rotational flexibility of the hydroxymethyl group allows
the relative probability for the $O6$--\ce{Na+} contacts to increase with
temperature and to become the largest at temperatures $>410$ K. In the bottom
part of Table~\ref{tab:1} we show the results for the relative three--body
contacts (joint probabilities) involving \ce{Na+}, $Ox$ and $Oy$ ($x,y=2,3,6$, $x \neq
y$) at $T=298$, $T=402$ and $T=502$ K. Full temperature dependence data show
that the \ce{Na+} coordination involving $O2$ and $O3$ atoms has, by far, the
highest statistical weight across the temperature interval considered in our
simulations.

In terms of intramolecular cellulose hydrogen bonding in both the
NaCl and pure water simulations of the solvated tetramer, we observe a dominant
$O3-O5$ bond and marginal contributions from $O6-O3$, $O3-O6$, $O6-O2$ and $O2-O6$
hydrogen bonds. The percentage contribution of the $O3-O5$ hydrogen bond
decreases linearly from $\sim 80-90\%$ to $\sim 57\%$ with temperature
increasing from $298$ K to $400$ K. Conversely, the same temperature increase
is associated with an increase in the percentage contribution of $O6-O3$,
$O3-O6$, $O6-O2$ and $O2-O6$ hydrogen bonds together from $7-9\%$ to $23-25\%$
(data not shown). 

\subsection{Cellulose I$_{\beta}$ Fibril\label{sec:res5}}
The comparison between the hydroxymethyl population in a model cellulose fibril
in NaCl aqueous solution (at $298$ and $400$ K) and an identical fibril in pure
water \cite{ShiGio2011} (at $298$ and $400$ K) is reported in Table~\ref{tab:1b}.
The results show that the presence of NaCl perturbs the hydroxymethyl rotational
state occupancy favoring the $gt$ and $gg$ conformations over $tg$ in both the
fibril crystalline core and its surface at both temperatures. At $400$ K, the
$gg$ rotational state becomes the dominant one both on the fibril surface and
in its crystalline core. Despite the change in the hydroxymethyl rotational
state occupancy, due to the presence of NaCl, the fibril maintains I$_{\beta}$
symmetry in its crystalline core at both $298$ and $400$ K (data not shown). 
We did not observe any non-chair--like conformation in our simulations of the
cellulose fibrils (neither at $298$ nor at $400$ K).

\subsection{Cellulose I$_{\beta}$ Fibril--\ce{Na+} Interaction\label{sec:res6}}
In \ref{tab:2} we report the results for the relevant two and three-body
relative contact probabilities involving \ce{Na+} and the cellulose oxygens
$Ox$ ($x=2,3,6$) in a model cellulose fibril. 
The cellulose-\ce{Na+} contacts analyzed in our study apply to the cellulose chains 
on the fibril surface and exposed directly to the solvent.
We did not observe any substantial penetration of the \ce{Na+} ions within the cellulose fibril. 
Single contacts are highly favored over bridging positions, and among those single contacts, $O6$ has
the highest probability of contact to \ce{Na+} at both $T=298$ and 
$T=400$ K. The contacts between the cellulose oxygens $O2,O3,O6$ and \ce{Na+}
account for over $70\%$ of the total contacts between cellulose heavy atoms and
\ce{Na+} \ce{Cl-} pairs. Specifically, the ratio between \ce{Cl-} and \ce{Na+} 
contacts with the cellulose heavy atoms is $0.29$ at $298$ K and $0.27$ at
$400$ K. $O2$ shows the highest probability of contact to \ce{Cl-} at $298$ K
and $400$ K. We did not observe any simultaneous contact between \ce{Cl-} and
cellulose oxygen pairs (bridging positions).

The internal hydrogen bond for the cellulose fibril in the presence of NaCl
shows some differences when compared with the results from simulations in pure
water.\cite{ShiGio2011} When the hydroxymethyl group deviates from the $tg$ 
conformation (typical of native crystalline cellulose) we expect the weakening
(or the disappearance) of the native intramolecular hydrogen bonds involving
the hydroxymethyl oxygen $O6$ and the appearance of new hydrogen bonds connecting 
neighboring layers within the crystal.\cite{Nishiyama2002,Wada2004} Indeed, a
comparison with previous simulations in pure water\cite{ShiGio2011} shows that 
the presence of NaCl leads to a lower number of intramolecular hydrogen bonds
(Fig. S3 and S4, top row) with the number decreasing with
increasing temperature from $298$ to $400$ K, and a higher number of intersheet hydrogen
bonds (Fig. S3 and S4, bottom row) with the number increasing  with
increasing temperature from $298$ to $400$ K. The intermolecular/intrasheet
hydrogen bond network (Fig. S3 and S4, center row) does not seem to be perturbed
by the presence of NaCl at $298$ K. Conversely, at $400$ K the presence of NaCl
results in a lower number of intermolecular/intrasheet hydrogen bonds.

In Table~\ref{tab:3} we show the relative contributions to the cellulose-water
hydrogen bond network at $298$ K of oxygens $O2,O3,O6$ for a model cellulose
fibril with NaCl (Fib-NaCl) and without NaCl (Fib).\cite{ShiGio2011} Our data
reveal that the presence of NaCl (even at relatively low concentrations - see
Methods Section) results in a decrease of the relative contribution of the
hydroxymethyl oxygen $O6$ possibly due to its high relative affinity to \ce{Na+}. 
Typically, the water coordination to $O3$ is the lowest due to presence of
persistent $O3-O5$ intrachain hydrogen bond. Interestingly, a slight increase is
seen in $O3$ hydrogen bonding to water in the presence of NaCl. The
cellulose-water hydrogen bonding maintains these same general trends also at
$400$ K in the crystalline cellulose simulations. Interestingly, for the
solvated tetramer neither the presence of NaCl nor a change in temperature (in
the interval $298-550$ K) alters the relative contributions of the oxygens
$O2$, $O3$ and $O6$ to the cellulose--water hydrogen bond network.

\section{Conclusion \label{sec:res8}}
Inorganic salts, in general and NaCl in particular, are known to adversely
affect biomass fast pyrolysis and to increase the efficiency of catalytic
cellulose degradation in the aqueous environment. The purpose of this
computational study was to resolve the atomistic details of the interactions of 
oligomeric and fibrillar forms of solvated cellulose with \ce{Na+}. 

Our classical MD simulations show that the dominant interaction is the
one between the \ce{Na+} and the hydroxyl oxygen atoms in cellulose. 

In more detail, we observe that, in cellulose, the two backbone oxygens
$O2$, $O3$ and the sidechain oxygen $O6$ account for most of the interactions
with \ce{Na+}, with the hydroxymethyl oxygen $O6$ dominating at high
temperatures ($>410$ K) in solvated cellulose and at both $298$ K and $400$ K
in crystalline cellulose. In solvated cellulose, we also noticed a high
affinity for \ce{Na+} to be in a ``bridging'' position between the $O2$ and
$O3$ backbone oxygens.

Our simulations also show only minor structural changes in the pyranose ring due 
to the presence of NaCl. 
The chair conformation was the dominant ring structure for both oligomeric and
crystalline cellulose in the classical MD simulations. In crystalline
cellulose, the presence of NaCl perturbs the rotational state population of the
hydroxymethyl group. As a consequence of the perturbation of the hydroxymethyl
rotational state population, the hydrogen bonding network in crystalline
cellulose is also perturbed by the presence of NaCl. In particular we observed
the disappearance of a number of intramolecular hydrogen bonds and a
consequential emergence of intersheet hydrogen bonds as seen previously in the
high temperature MD simulations of native cellulose and cellulose III$_{I}$ \cite{ShiGio2011,MatthewsHT2011}. 
The enhanced understanding of the interactions of sodium chloride with cellulose
gained from this computational study will provide valuable information for the
design of cost-effective thermochemical degradation protocols for cellulosic
biomass.

\section{Acknowledgements}
This work was supported by CNLS and LANL Institutional Computing. 

\bibliography{tutto}

\providecommand{\noopsort}[1]{}\providecommand{\singleletter}[1]{#1}%
\begin{thebibliography}{28}%
\makeatletter
\providecommand \@ifxundefined [1]{%
 \@ifx{#1\undefined}
}%
\providecommand \@ifnum [1]{%
 \ifnum #1\expandafter \@firstoftwo
 \else \expandafter \@secondoftwo
 \fi
}%
\providecommand \@ifx [1]{%
 \ifx #1\expandafter \@firstoftwo
 \else \expandafter \@secondoftwo
 \fi
}%
\providecommand \natexlab [1]{#1}%
\providecommand \enquote  [1]{``#1''}%
\providecommand \bibnamefont  [1]{#1}%
\providecommand \bibfnamefont [1]{#1}%
\providecommand \citenamefont [1]{#1}%
\providecommand \href@noop [0]{\@secondoftwo}%
\providecommand \href [0]{\begingroup \@sanitize@url \@href}%
\providecommand \@href[1]{\@@startlink{#1}\@@href}%
\providecommand \@@href[1]{\endgroup#1\@@endlink}%
\providecommand \@sanitize@url [0]{\catcode `\\12\catcode `\$12\catcode
  `\&12\catcode `\#12\catcode `\^12\catcode `\_12\catcode `\%12\relax}%
\providecommand \@@startlink[1]{}%
\providecommand \@@endlink[0]{}%
\providecommand \url  [0]{\begingroup\@sanitize@url \@url }%
\providecommand \@url [1]{\endgroup\@href {#1}{\urlprefix }}%
\providecommand \urlprefix  [0]{URL }%
\providecommand \Eprint [0]{\href }%
\@ifxundefined \urlstyle {%
  \providecommand \doi  [0]{\begingroup \@sanitize@url \@doi}%
  \providecommand \@doi [1]{\endgroup \@@startlink {\doibase
  #1}doi:\discretionary {}{}{}#1\@@endlink }%
}{%
  \providecommand \doi  [0]{doi:\discretionary{}{}{}\begingroup
  \urlstyle{rm}\Url }%
}%
\providecommand \doibase [0]{http://dx.doi.org/}%
\providecommand \Doi [0]{\begingroup \@sanitize@url \@Doi }%
\providecommand \@Doi  [1]{\endgroup\@@startlink{\doibase#1}\@@Doi}%
\providecommand \@@Doi [1]{#1\@@endlink}%
\providecommand \selectlanguage [0]{\@gobble}%
\providecommand \bibinfo  [0]{\@secondoftwo}%
\providecommand \bibfield  [0]{\@secondoftwo}%
\providecommand \translation [1]{[#1]}%
\providecommand \BibitemOpen [0]{}%
\providecommand \bibitemStop [0]{}%
\providecommand \bibitemNoStop [0]{.\EOS\space}%
\providecommand \EOS [0]{\spacefactor3000\relax}%
\providecommand \BibitemShut  [1]{\csname bibitem#1\endcsname}%
\bibitem [{\citenamefont {Mohan}\ \emph {et~al.}(2006)\citenamefont {Mohan},
  \citenamefont {Pittman},\ and\ \citenamefont {Steele}}]{Mohan2006}%
  \BibitemOpen
  \bibfield  {author} {\bibinfo {author} {\bibfnamefont {D.}~\bibnamefont
  {Mohan}}, \bibinfo {author} {\bibfnamefont {C.~U.}\ \bibnamefont {Pittman},
  \bibfnamefont {Jr.}}, \ and\ \bibinfo {author} {\bibfnamefont {P.~H.}\
  \bibnamefont {Steele}},\ }\bibfield  {title} {\enquote {\bibinfo {title}
  {{Pyrolysis of Wood/Biomass for Bio-oil: A Critical Review}},}\ }\Doi
  {10.1021/ef0502397} {\bibfield  {journal} {\bibinfo  {journal} {Energy
  Fuels},\ }\textbf {\bibinfo {volume} {20}},\ \bibinfo {pages} {848} (\bibinfo
  {year} {2006})},\ ISSN \bibinfo {issn} {0887-0624}\BibitemShut {NoStop}%
\bibitem [{\citenamefont {Patwardhan}\ \emph {et~al.}(2009)\citenamefont
  {Patwardhan}, \citenamefont {Satrio}, \citenamefont {Brown},\ and\
  \citenamefont {Shanks}}]{Patwardhan2009}%
  \BibitemOpen
  \bibfield  {author} {\bibinfo {author} {\bibfnamefont {P.~R.}\ \bibnamefont
  {Patwardhan}}, \bibinfo {author} {\bibfnamefont {J.~A.}\ \bibnamefont
  {Satrio}}, \bibinfo {author} {\bibfnamefont {R.~C.}\ \bibnamefont {Brown}}, \
  and\ \bibinfo {author} {\bibfnamefont {B.~H.}\ \bibnamefont {Shanks}},\
  }\bibfield  {title} {\enquote {\bibinfo {title} {{Product Distribution from
  Fast Pyrolysis of Glucose-Based Carbohydrates}},}\ }\Doi
  {10.1016/j.jaap.2009.08.007} {\bibfield  {journal} {\bibinfo  {journal} {J.
  Anal. Appl. Pyrolysis},\ }\textbf {\bibinfo {volume} {86}},\ \bibinfo {pages}
  {323} (\bibinfo {year} {2009})},\ ISSN \bibinfo {issn} {01652370}\BibitemShut
  {NoStop}%
\bibitem [{\citenamefont {Patwardhan}\ \emph {et~al.}(2011)\citenamefont
  {Patwardhan}, \citenamefont {Brown},\ and\ \citenamefont
  {Shanks}}]{Patwardhan2011}%
  \BibitemOpen
  \bibfield  {author} {\bibinfo {author} {\bibfnamefont {P.~R.}\ \bibnamefont
  {Patwardhan}}, \bibinfo {author} {\bibfnamefont {R.~C.}\ \bibnamefont
  {Brown}}, \ and\ \bibinfo {author} {\bibfnamefont {B.~H.}\ \bibnamefont
  {Shanks}},\ }\bibfield  {title} {\enquote {\bibinfo {title} {{Product
  Distribution from the Fast Pyrolysis of Hemicellulose.}}}\ }\Doi
  {10.1002/cssc.201000425} {\bibfield  {journal} {\bibinfo  {journal}
  {ChemSusChem},\ }\textbf {\bibinfo {volume} {4}},\ \bibinfo {pages} {636}
  (\bibinfo {year} {2011})},\ ISSN \bibinfo {issn} {1864-564X}\BibitemShut
  {NoStop}%
\bibitem [{\citenamefont {Scott}\ \emph {et~al.}(2001)\citenamefont {Scott},
  \citenamefont {Paterson}, \citenamefont {Piskorz},\ and\ \citenamefont
  {Radlein}}]{Scott2001}%
  \BibitemOpen
  \bibfield  {author} {\bibinfo {author} {\bibfnamefont {D.~S.}\ \bibnamefont
  {Scott}}, \bibinfo {author} {\bibfnamefont {L.}~\bibnamefont {Paterson}},
  \bibinfo {author} {\bibfnamefont {J.}~\bibnamefont {Piskorz}}, \ and\
  \bibinfo {author} {\bibfnamefont {D.}~\bibnamefont {Radlein}},\ }\bibfield
  {title} {\enquote {\bibinfo {title} {{Pretreatment of Poplar Wood for Fast
  Pyrolysis: Rate of Cation Removal}},}\ }\Doi {10.1016/S0165-2370(00)00108-X}
  {\bibfield  {journal} {\bibinfo  {journal} {J. Anal. Appl. Pyrolysis},\
  }\textbf {\bibinfo {volume} {57}},\ \bibinfo {pages} {169} (\bibinfo {year}
  {2001})}\BibitemShut {NoStop}%
\bibitem [{\citenamefont {Patwardhan}\ \emph {et~al.}(2010)\citenamefont
  {Patwardhan}, \citenamefont {Satrio}, \citenamefont {Brown},\ and\
  \citenamefont {Shanks}}]{Patwardhan2010}%
  \BibitemOpen
  \bibfield  {author} {\bibinfo {author} {\bibfnamefont {P.~R.}\ \bibnamefont
  {Patwardhan}}, \bibinfo {author} {\bibfnamefont {J.~A.}\ \bibnamefont
  {Satrio}}, \bibinfo {author} {\bibfnamefont {R.~C.}\ \bibnamefont {Brown}}, \
  and\ \bibinfo {author} {\bibfnamefont {B.~H.}\ \bibnamefont {Shanks}},\
  }\bibfield  {title} {\enquote {\bibinfo {title} {Influence of inorganic salts
  on the primary pyrolysis products of cellulose},}\ }\href@noop {} {\bibfield
  {journal} {\bibinfo  {journal} {Bioresour. Technol.},\ }\textbf {\bibinfo
  {volume} {101}},\ \bibinfo {pages} {4646} (\bibinfo {year}
  {2010})}\BibitemShut {NoStop}%
\bibitem [{\citenamefont {Encinar}\ \emph {et~al.}(2001)\citenamefont
  {Encinar}, \citenamefont {Gonz\'{a}lez}, \citenamefont {Rodr\'{\i}guez},\
  and\ \citenamefont {Ramiro}}]{Encinar2001}%
  \BibitemOpen
  \bibfield  {author} {\bibinfo {author} {\bibfnamefont {J.~M.}\ \bibnamefont
  {Encinar}}, \bibinfo {author} {\bibfnamefont {J.~F.}\ \bibnamefont
  {Gonz\'{a}lez}}, \bibinfo {author} {\bibfnamefont {J.~J.}\ \bibnamefont
  {Rodr\'{\i}guez}}, \ and\ \bibinfo {author} {\bibfnamefont {M.~J.}\
  \bibnamefont {Ramiro}},\ }\bibfield  {title} {\enquote {\bibinfo {title}
  {{Catalysed and uncatalysed steam gasification of eucalyptus char: influence
  of variables and kinetic study}},}\ }\href@noop {} {\bibfield  {journal}
  {\bibinfo  {journal} {Fuel},\ }\textbf {\bibinfo {volume} {80}},\ \bibinfo
  {pages} {2025} (\bibinfo {year} {2001})}\BibitemShut {NoStop}%
\bibitem [{\citenamefont {vom Stein}\ \emph {et~al.}(2010)\citenamefont {vom
  Stein}, \citenamefont {Grande}, \citenamefont {Sibilla}, \citenamefont
  {Commandeur}, \citenamefont {Fischer}, \citenamefont {Leitner},\ and\
  \citenamefont {Dominguez~de Maria}}]{vomStein2010}%
  \BibitemOpen
  \bibfield  {author} {\bibinfo {author} {\bibfnamefont {T.}~\bibnamefont {vom
  Stein}}, \bibinfo {author} {\bibfnamefont {P.}~\bibnamefont {Grande}},
  \bibinfo {author} {\bibfnamefont {F.}~\bibnamefont {Sibilla}}, \bibinfo
  {author} {\bibfnamefont {U.}~\bibnamefont {Commandeur}}, \bibinfo {author}
  {\bibfnamefont {R.}~\bibnamefont {Fischer}}, \bibinfo {author} {\bibfnamefont
  {W.}~\bibnamefont {Leitner}}, \ and\ \bibinfo {author} {\bibfnamefont
  {P.}~\bibnamefont {Dominguez~de Maria}},\ }\bibfield  {title} {\enquote
  {\bibinfo {title} {Salt-assisted organic-acid-catalyzed depolymerization of
  cellulose},}\ }\href@noop {} {\bibfield  {journal} {\bibinfo  {journal}
  {Green Chem.},\ }\textbf {\bibinfo {volume} {12}},\ \bibinfo {pages} {1844}
  (\bibinfo {year} {2010})}\BibitemShut {NoStop}%
\bibitem [{\citenamefont {Potvin}\ \emph {et~al.}(2011)\citenamefont {Potvin},
  \citenamefont {Sorlien}, \citenamefont {Hegner}, \citenamefont {DeBoef},\
  and\ \citenamefont {Lucht}}]{Potvin2011}%
  \BibitemOpen
  \bibfield  {author} {\bibinfo {author} {\bibfnamefont {J.}~\bibnamefont
  {Potvin}}, \bibinfo {author} {\bibfnamefont {E.}~\bibnamefont {Sorlien}},
  \bibinfo {author} {\bibfnamefont {J.}~\bibnamefont {Hegner}}, \bibinfo
  {author} {\bibfnamefont {B.}~\bibnamefont {DeBoef}}, \ and\ \bibinfo {author}
  {\bibfnamefont {B.~L.}\ \bibnamefont {Lucht}},\ }\bibfield  {title} {\enquote
  {\bibinfo {title} {Effect of nacl on the conversion of cellulose to glucose
  and levulinic acid},}\ }\href@noop {} {\bibfield  {journal} {\bibinfo
  {journal} {Tetrahedron Lett.},\ }\textbf {\bibinfo {volume} {52}},\ \bibinfo
  {pages} {5891} (\bibinfo {year} {2011})}\BibitemShut {NoStop}%
\bibitem [{\citenamefont {Swensen}\ and\ \citenamefont
  {Wang}(1986)}]{Swensen1986}%
  \BibitemOpen
  \bibfield  {author} {\bibinfo {author} {\bibfnamefont {R.}~\bibnamefont
  {Swensen}}\ and\ \bibinfo {author} {\bibfnamefont {J.}~\bibnamefont {Wang}},\
  }\bibfield  {title} {\enquote {\bibinfo {title} {Replica monte carlo
  simulation of spin glasses},}\ }\href@noop {} {\bibfield  {journal} {\bibinfo
   {journal} {Phys. Rev. Lett.},\ }\textbf {\bibinfo {volume} {57}},\ \bibinfo
  {pages} {2607} (\bibinfo {year} {1986})}\BibitemShut {NoStop}%
\bibitem [{\citenamefont {Sugita}\ and\ \citenamefont
  {Okamoto}(1999)}]{Sugita1999}%
  \BibitemOpen
  \bibfield  {author} {\bibinfo {author} {\bibfnamefont {Y.}~\bibnamefont
  {Sugita}}\ and\ \bibinfo {author} {\bibfnamefont {Y.}~\bibnamefont
  {Okamoto}},\ }\bibfield  {title} {\enquote {\bibinfo {title} {Replica
  exchange molecular dynamics method for protein folding},}\ }\href@noop {}
  {\bibfield  {journal} {\bibinfo  {journal} {Chem. Phys. Lett.},\ }\textbf
  {\bibinfo {volume} {314}},\ \bibinfo {pages} {141} (\bibinfo {year}
  {1999})}\BibitemShut {NoStop}%
\bibitem [{gly()}]{glycam-url}%
  \BibitemOpen
  \href {http://glycam.ccrc.uga.edu/AMBER/index.html} {}\bibinfo {howpublished}
  {http://glycam.ccrc.uga.edu/AMBER/index.html}\BibitemShut {NoStop}%
\bibitem [{\citenamefont {Shen}\ \emph {et~al.}(2009)\citenamefont {Shen},
  \citenamefont {Langan}, \citenamefont {French}, \citenamefont {Johnson},\
  and\ \citenamefont {Gnanakaran}}]{Shen2009}%
  \BibitemOpen
  \bibfield  {author} {\bibinfo {author} {\bibfnamefont {T.}~\bibnamefont
  {Shen}}, \bibinfo {author} {\bibfnamefont {P.}~\bibnamefont {Langan}},
  \bibinfo {author} {\bibfnamefont {A.~D.}\ \bibnamefont {French}}, \bibinfo
  {author} {\bibfnamefont {G.~P.}\ \bibnamefont {Johnson}}, \ and\ \bibinfo
  {author} {\bibfnamefont {S.}~\bibnamefont {Gnanakaran}},\ }\bibfield  {title}
  {\enquote {\bibinfo {title} {Conformational flexibility of soluble cellulose
  oligomers: chain length and temperature dependence.}}\ }\Doi
  {10.1021/ja9034158} {\bibfield  {journal} {\bibinfo  {journal} {J. Am. Chem.
  Soc.},\ }\textbf {\bibinfo {volume} {131}},\ \bibinfo {pages} {14786}
  (\bibinfo {year} {2009})}\BibitemShut {NoStop}%
\bibitem [{\citenamefont {Bellesia}\ and\ \citenamefont
  {Shea}(2009)}]{GioBJ2009}%
  \BibitemOpen
  \bibfield  {author} {\bibinfo {author} {\bibfnamefont {G.}~\bibnamefont
  {Bellesia}}\ and\ \bibinfo {author} {\bibfnamefont {J.-E.}\ \bibnamefont
  {Shea}},\ }\bibfield  {title} {\enquote {\bibinfo {title} {What determines
  the structure and stability of kffe monomers, dimers and protofibrils},}\
  }\href@noop {} {\bibfield  {journal} {\bibinfo  {journal} {Biophys. J.},\
  }\textbf {\bibinfo {volume} {96}},\ \bibinfo {pages} {875} (\bibinfo {year}
  {2009})}\BibitemShut {NoStop}%
\bibitem [{\citenamefont {Chundawat}\ \emph {et~al.}(2011)\citenamefont
  {Chundawat}, \citenamefont {Bellesia}, \citenamefont {Uppugundla},
  \citenamefont {da~Costa~Sousa}, \citenamefont {Gao}, \citenamefont {Cheh},
  \citenamefont {Agarwal}, \citenamefont {Bianchetti}, \citenamefont
  {Phillips}, \citenamefont {Langan}, \citenamefont {Balan}, \citenamefont
  {Gnanakaran},\ and\ \citenamefont {Dale}}]{ShiGio2011}%
  \BibitemOpen
  \bibfield  {author} {\bibinfo {author} {\bibfnamefont {S.~P.}\ \bibnamefont
  {Chundawat}}, \bibinfo {author} {\bibfnamefont {G.}~\bibnamefont {Bellesia}},
  \bibinfo {author} {\bibfnamefont {N.}~\bibnamefont {Uppugundla}}, \bibinfo
  {author} {\bibfnamefont {L.}~\bibnamefont {da~Costa~Sousa}}, \bibinfo
  {author} {\bibfnamefont {D.}~\bibnamefont {Gao}}, \bibinfo {author}
  {\bibfnamefont {A.}~\bibnamefont {Cheh}}, \bibinfo {author} {\bibfnamefont
  {U.~P.}\ \bibnamefont {Agarwal}}, \bibinfo {author} {\bibfnamefont {C.~M.}\
  \bibnamefont {Bianchetti}}, \bibinfo {author} {\bibfnamefont {G.~N.~J.}\
  \bibnamefont {Phillips}}, \bibinfo {author} {\bibfnamefont {P.}~\bibnamefont
  {Langan}}, \bibinfo {author} {\bibfnamefont {V.}~\bibnamefont {Balan}},
  \bibinfo {author} {\bibfnamefont {S.}~\bibnamefont {Gnanakaran}}, \ and\
  \bibinfo {author} {\bibfnamefont {B.~E.}\ \bibnamefont {Dale}},\ }\bibfield
  {title} {\enquote {\bibinfo {title} {Restructuring crystalline cellulose
  hydrogen bond network enhances its depolymerization rate},}\ }\href@noop {}
  {\bibfield  {journal} {\bibinfo  {journal} {J. Am. Chem. Soc.},\ }\textbf
  {\bibinfo {volume} {133}},\ \bibinfo {pages} {11163} (\bibinfo {year}
  {2011})}\BibitemShut {NoStop}%
\bibitem [{\citenamefont {Wang}\ \emph {et~al.}(2000)\citenamefont {Wang},
  \citenamefont {Cieplak},\ and\ \citenamefont {Kollman}}]{Amber99}%
  \BibitemOpen
  \bibfield  {author} {\bibinfo {author} {\bibfnamefont {J.}~\bibnamefont
  {Wang}}, \bibinfo {author} {\bibfnamefont {P.}~\bibnamefont {Cieplak}}, \
  and\ \bibinfo {author} {\bibfnamefont {P.}~\bibnamefont {Kollman}},\
  }\bibfield  {title} {\enquote {\bibinfo {title} {How well does a restrained
  electrostatic potential (resp) model perform in calculating energies of
  organic and biological molecules ?}}\ }\href@noop {} {\bibfield  {journal}
  {\bibinfo  {journal} {J. Comput. Chem.},\ }\textbf {\bibinfo {volume} {21}},\
  \bibinfo {pages} {1049} (\bibinfo {year} {2000})}\BibitemShut {NoStop}%
\bibitem [{\citenamefont {Chen}\ and\ \citenamefont {Pappu}(2007)}]{Pappu2007}%
  \BibitemOpen
  \bibfield  {author} {\bibinfo {author} {\bibfnamefont {A.~A.}\ \bibnamefont
  {Chen}}\ and\ \bibinfo {author} {\bibfnamefont {R.~V.}\ \bibnamefont
  {Pappu}},\ }\bibfield  {title} {\enquote {\bibinfo {title} {Parameters of
  monovalent ions in the amber99 forcefield: assessment of inaccuracies and
  proposed improvements},}\ }\href@noop {} {\bibfield  {journal} {\bibinfo
  {journal} {J. Phys. Chem. B},\ }\textbf {\bibinfo {volume} {111}},\ \bibinfo
  {pages} {11884} (\bibinfo {year} {2007})}\BibitemShut {NoStop}%
\bibitem [{\citenamefont {Ryckaert}\ \emph {et~al.}(1977)\citenamefont
  {Ryckaert}, \citenamefont {Ciccotti},\ and\ \citenamefont
  {Berendsen}}]{Ryckaert1977}%
  \BibitemOpen
  \bibfield  {author} {\bibinfo {author} {\bibfnamefont {J.}~\bibnamefont
  {Ryckaert}}, \bibinfo {author} {\bibfnamefont {G.}~\bibnamefont {Ciccotti}},
  \ and\ \bibinfo {author} {\bibfnamefont {H.}~\bibnamefont {Berendsen}},\
  }\bibfield  {title} {\enquote {\bibinfo {title} {Numerical integration of the
  cartesian equations of motion of a system with constraints: Molecular
  dynamics of n-alkanes},}\ }\href@noop {} {\bibfield  {journal} {\bibinfo
  {journal} {J. Comput. Phys.},\ }\textbf {\bibinfo {volume} {23}},\ \bibinfo
  {pages} {327} (\bibinfo {year} {1977})}\BibitemShut {NoStop}%
\bibitem [{\citenamefont {Phillips}\ \emph {et~al.}(2005)\citenamefont
  {Phillips}, \citenamefont {Braun}, \citenamefont {Wang}, \citenamefont
  {Gumbart}, \citenamefont {Tajkhorshid}, \citenamefont {Villa}, \citenamefont
  {Chipot}, \citenamefont {Skeel}, \citenamefont {Kale},\ and\ \citenamefont
  {Schulten}}]{Phillips2005}%
  \BibitemOpen
  \bibfield  {author} {\bibinfo {author} {\bibfnamefont {J.}~\bibnamefont
  {Phillips}}, \bibinfo {author} {\bibfnamefont {R.}~\bibnamefont {Braun}},
  \bibinfo {author} {\bibfnamefont {W.}~\bibnamefont {Wang}}, \bibinfo {author}
  {\bibfnamefont {J.}~\bibnamefont {Gumbart}}, \bibinfo {author} {\bibfnamefont
  {E.}~\bibnamefont {Tajkhorshid}}, \bibinfo {author} {\bibfnamefont
  {E.}~\bibnamefont {Villa}}, \bibinfo {author} {\bibfnamefont
  {C.}~\bibnamefont {Chipot}}, \bibinfo {author} {\bibfnamefont
  {R.}~\bibnamefont {Skeel}}, \bibinfo {author} {\bibfnamefont
  {L.}~\bibnamefont {Kale}}, \ and\ \bibinfo {author} {\bibfnamefont
  {K.}~\bibnamefont {Schulten}},\ }\bibfield  {title} {\enquote {\bibinfo
  {title} {Scalable molecular dynamics with namd},}\ }\Doi {DOI
  10.1002/jcc.20289} {\bibfield  {journal} {\bibinfo  {journal} {J. Comput.
  Chem.},\ }\textbf {\bibinfo {volume} {26}},\ \bibinfo {pages} {1781}
  (\bibinfo {year} {2005})}\BibitemShut {NoStop}%
\bibitem [{\citenamefont {Kirschner}\ \emph {et~al.}(2008)\citenamefont
  {Kirschner}, \citenamefont {Yongye}, \citenamefont {Tschampel}, \citenamefont
  {Gonz{\'a}lez-Outeiri{\~n}o}, \citenamefont {Daniels}, \citenamefont
  {Foley},\ and\ \citenamefont {Woods}}]{Kirschner2008}%
  \BibitemOpen
  \bibfield  {author} {\bibinfo {author} {\bibfnamefont {K.}~\bibnamefont
  {Kirschner}}, \bibinfo {author} {\bibfnamefont {A.}~\bibnamefont {Yongye}},
  \bibinfo {author} {\bibfnamefont {S.}~\bibnamefont {Tschampel}}, \bibinfo
  {author} {\bibfnamefont {J.}~\bibnamefont {Gonz{\'a}lez-Outeiri{\~n}o}},
  \bibinfo {author} {\bibfnamefont {C.}~\bibnamefont {Daniels}}, \bibinfo
  {author} {\bibfnamefont {L.}~\bibnamefont {Foley}}, \ and\ \bibinfo {author}
  {\bibfnamefont {R.}~\bibnamefont {Woods}},\ }\bibfield  {title} {\enquote
  {\bibinfo {title} {Glycam06: A generalizable biomolecular force field.
  carbohydrates},}\ }\Doi {10.1002/jcc.20820} {\bibfield  {journal} {\bibinfo
  {journal} {J. Comput. Chem.},\ }\textbf {\bibinfo {volume} {29}},\ \bibinfo
  {pages} {622} (\bibinfo {year} {2008})},\ ISSN \bibinfo {issn}
  {1096-987X}\BibitemShut {NoStop}%
\bibitem [{\citenamefont {Jorgensen}\ \emph {et~al.}(1983)\citenamefont
  {Jorgensen}, \citenamefont {Chandrasekhar}, \citenamefont {D}, \citenamefont
  {Impey},\ and\ \citenamefont {L}}]{tip3p}%
  \BibitemOpen
  \bibfield  {author} {\bibinfo {author} {\bibfnamefont {W.~L.}\ \bibnamefont
  {Jorgensen}}, \bibinfo {author} {\bibfnamefont {J.}~\bibnamefont
  {Chandrasekhar}}, \bibinfo {author} {\bibfnamefont {M.~J.}\ \bibnamefont
  {D}}, \bibinfo {author} {\bibfnamefont {R.~W.}\ \bibnamefont {Impey}}, \ and\
  \bibinfo {author} {\bibfnamefont {K.~M.}\ \bibnamefont {L}},\ }\bibfield
  {title} {\enquote {\bibinfo {title} {Comparison of simple potential functions
  for simulating liquid water},}\ }\href@noop {} {\bibfield  {journal}
  {\bibinfo  {journal} {J. Chem. Phys.},\ }\textbf {\bibinfo {volume} {79}},\
  \bibinfo {pages} {926} (\bibinfo {year} {1983})}\BibitemShut {NoStop}%
\bibitem [{\citenamefont {Bellesia}\ \emph {et~al.}(2011)\citenamefont
  {Bellesia}, \citenamefont {Chundawat}, \citenamefont {Langan}, \citenamefont
  {Dale},\ and\ \citenamefont {Gnanakaran}}]{Gioamm2011}%
  \BibitemOpen
  \bibfield  {author} {\bibinfo {author} {\bibfnamefont {G.}~\bibnamefont
  {Bellesia}}, \bibinfo {author} {\bibfnamefont {S.~P.~S.}\ \bibnamefont
  {Chundawat}}, \bibinfo {author} {\bibfnamefont {P.}~\bibnamefont {Langan}},
  \bibinfo {author} {\bibfnamefont {B.~E.}\ \bibnamefont {Dale}}, \ and\
  \bibinfo {author} {\bibfnamefont {S.}~\bibnamefont {Gnanakaran}},\ }\bibfield
   {title} {\enquote {\bibinfo {title} {Probing the early events associated
  with liquid ammonia pretreatment of native crystalline cellulose},}\
  }\href@noop {} {\bibfield  {journal} {\bibinfo  {journal} {Journal of
  Physical Chemistry B},\ }\textbf {\bibinfo {volume} {115}},\ \bibinfo {pages}
  {9782} (\bibinfo {year} {2011})}\BibitemShut {NoStop}%
\bibitem [{\citenamefont {Matthews}\ \emph {et~al.}(2012)\citenamefont
  {Matthews}, \citenamefont {Beckham}, \citenamefont
  {Bergenstr{\aa}hle-Wohlert}, \citenamefont {Brady}, \citenamefont {Himmel},\
  and\ \citenamefont {Crowley}}]{Matthews2012comparison}%
  \BibitemOpen
  \bibfield  {author} {\bibinfo {author} {\bibfnamefont {J.}~\bibnamefont
  {Matthews}}, \bibinfo {author} {\bibfnamefont {G.}~\bibnamefont {Beckham}},
  \bibinfo {author} {\bibfnamefont {M.}~\bibnamefont
  {Bergenstr{\aa}hle-Wohlert}}, \bibinfo {author} {\bibfnamefont
  {J.}~\bibnamefont {Brady}}, \bibinfo {author} {\bibfnamefont
  {M.}~\bibnamefont {Himmel}}, \ and\ \bibinfo {author} {\bibfnamefont
  {M.}~\bibnamefont {Crowley}},\ }\bibfield  {title} {\enquote {\bibinfo
  {title} {Comparison of cellulose i$\beta$ simulations with three carbohydrate
  force fields},}\ }\href@noop {} {\bibfield  {journal} {\bibinfo  {journal}
  {Journal of Chemical Theory and Computation},\ }\textbf {\bibinfo {volume}
  {8}},\ \bibinfo {pages} {735} (\bibinfo {year} {2012})}\BibitemShut {NoStop}%
\bibitem [{\citenamefont {Feller}\ \emph {et~al.}(1995)\citenamefont {Feller},
  \citenamefont {Zhang}, \citenamefont {Pastor},\ and\ \citenamefont
  {Brooks}}]{Feller1995}%
  \BibitemOpen
  \bibfield  {author} {\bibinfo {author} {\bibfnamefont {S.}~\bibnamefont
  {Feller}}, \bibinfo {author} {\bibfnamefont {Y.}~\bibnamefont {Zhang}},
  \bibinfo {author} {\bibfnamefont {R.}~\bibnamefont {Pastor}}, \ and\ \bibinfo
  {author} {\bibfnamefont {B.}~\bibnamefont {Brooks}},\ }\bibfield  {title}
  {\enquote {\bibinfo {title} {Constant pressure molecular dynamics simulation:
  The langevin piston method},}\ }\href@noop {} {\bibfield  {journal} {\bibinfo
   {journal} {J. Chem. Phys.},\ }\textbf {\bibinfo {volume} {103}},\ \bibinfo
  {pages} {4613} (\bibinfo {year} {1995})}\BibitemShut {NoStop}%
\bibitem [{\citenamefont {Martyna}\ \emph {et~al.}(1994)\citenamefont
  {Martyna}, \citenamefont {Tobias},\ and\ \citenamefont {L}}]{Martyna1994}%
  \BibitemOpen
  \bibfield  {author} {\bibinfo {author} {\bibfnamefont {G.~J.}\ \bibnamefont
  {Martyna}}, \bibinfo {author} {\bibfnamefont {D.~J.}\ \bibnamefont {Tobias}},
  \ and\ \bibinfo {author} {\bibfnamefont {K.~M.}\ \bibnamefont {L}},\
  }\bibfield  {title} {\enquote {\bibinfo {title} {Constant pressure molecular
  dynamics algorithms},}\ }\href@noop {} {\bibfield  {journal} {\bibinfo
  {journal} {J. Chem. Phys.},\ }\textbf {\bibinfo {volume} {101}},\ \bibinfo
  {pages} {4177} (\bibinfo {year} {1994})}\BibitemShut {NoStop}%
\bibitem [{\citenamefont {Darden}\ \emph {et~al.}(1993)\citenamefont {Darden},
  \citenamefont {York},\ and\ \citenamefont {Pedersen}}]{Darden1993}%
  \BibitemOpen
  \bibfield  {author} {\bibinfo {author} {\bibfnamefont {T.}~\bibnamefont
  {Darden}}, \bibinfo {author} {\bibfnamefont {D.}~\bibnamefont {York}}, \ and\
  \bibinfo {author} {\bibfnamefont {L.}~\bibnamefont {Pedersen}},\ }\bibfield
  {title} {\enquote {\bibinfo {title} {Particle mesh ewald: An n⋅log(n)
  method for ewald sums in large systems},}\ }\href@noop {} {\bibfield
  {journal} {\bibinfo  {journal} {J. Chem. Phys.},\ }\textbf {\bibinfo {volume}
  {98}},\ \bibinfo {pages} {10089} (\bibinfo {year} {1993})}\BibitemShut
  {NoStop}%
\bibitem [{\citenamefont {Nishiyama}\ \emph {et~al.}(2002)\citenamefont
  {Nishiyama}, \citenamefont {Langan},\ and\ \citenamefont
  {Chanzy}}]{Nishiyama2002}%
  \BibitemOpen
  \bibfield  {author} {\bibinfo {author} {\bibfnamefont {Y.}~\bibnamefont
  {Nishiyama}}, \bibinfo {author} {\bibfnamefont {P.}~\bibnamefont {Langan}}, \
  and\ \bibinfo {author} {\bibfnamefont {H.}~\bibnamefont {Chanzy}},\
  }\bibfield  {title} {\enquote {\bibinfo {title} {Crystal structure and
  hydrogen-bonding system in cellulose ibeta from synchrotron x-ray and neutron
  fiber diffraction.}}\ }\href@noop {} {\bibfield  {journal} {\bibinfo
  {journal} {J. Am. Chem. Soc.},\ }\textbf {\bibinfo {volume} {124}},\ \bibinfo
  {pages} {9074} (\bibinfo {year} {2002})}\BibitemShut {NoStop}%
\bibitem [{\citenamefont {Wada}\ \emph {et~al.}(2004)\citenamefont {Wada},
  \citenamefont {Chanzy}, \citenamefont {Nishiyama},\ and\ \citenamefont
  {Langan}}]{Wada2004}%
  \BibitemOpen
  \bibfield  {author} {\bibinfo {author} {\bibfnamefont {M.}~\bibnamefont
  {Wada}}, \bibinfo {author} {\bibfnamefont {H.}~\bibnamefont {Chanzy}},
  \bibinfo {author} {\bibfnamefont {Y.}~\bibnamefont {Nishiyama}}, \ and\
  \bibinfo {author} {\bibfnamefont {P.}~\bibnamefont {Langan}},\ }\bibfield
  {title} {\enquote {\bibinfo {title} {Cellulose iii(i) crystal structure and
  hydrogen bonding by synchrotron x-ray and neutron fiber diffracion},}\
  }\href@noop {} {\bibfield  {journal} {\bibinfo  {journal} {Macromolecules},\
  }\textbf {\bibinfo {volume} {37}},\ \bibinfo {pages} {8548} (\bibinfo {year}
  {2004})}\BibitemShut {NoStop}%
\bibitem [{\citenamefont {Matthews}\ \emph {et~al.}(2011)\citenamefont
  {Matthews}, \citenamefont {Bergenstr{\aa}hle}, \citenamefont {Beckham},
  \citenamefont {Himmel}, \citenamefont {Nimlos}, \citenamefont {Brady},\ and\
  \citenamefont {Crowley}}]{MatthewsHT2011}%
  \BibitemOpen
  \bibfield  {author} {\bibinfo {author} {\bibfnamefont {J.~F.}\ \bibnamefont
  {Matthews}}, \bibinfo {author} {\bibfnamefont {M.}~\bibnamefont
  {Bergenstr{\aa}hle}}, \bibinfo {author} {\bibfnamefont {G.~T.}\ \bibnamefont
  {Beckham}}, \bibinfo {author} {\bibfnamefont {M.~E.}\ \bibnamefont {Himmel}},
  \bibinfo {author} {\bibfnamefont {M.~R.}\ \bibnamefont {Nimlos}}, \bibinfo
  {author} {\bibfnamefont {J.~W.}\ \bibnamefont {Brady}}, \ and\ \bibinfo
  {author} {\bibfnamefont {M.~F.}\ \bibnamefont {Crowley}},\ }\bibfield
  {title} {\enquote {\bibinfo {title} {High-temperature behavior of cellulose
  i},}\ }\Doi {10.1021/jp1106839} {\bibfield  {journal} {\bibinfo  {journal}
  {The Journal of Physical Chemistry B},\ }\textbf {\bibinfo {volume} {115}},\
  \bibinfo {pages} {2155} (\bibinfo {year} {2011})},\ \Eprint
  {http://arxiv.org/abs/http://pubs.acs.org/doi/pdf/10.1021/jp1106839}
  {http://pubs.acs.org/doi/pdf/10.1021/jp1106839} \BibitemShut {NoStop}%
\end{thebibliography}%

\newpage

\begin{table}[h]
\caption{\label{tab:1}
Cellulose tetramer in NaCl aqueous solution.
Two and three-body relative contact probabilities between \ce{Na+} and cellulose
oxygens $Ox$ where $x=2,3,6$. For example, the two--body relative contact
probability between \ce{Na+} ions and $O6$ oxygens is calculated as
$P(Na-O6)=Na-O6-contacts/(Na-O2-contacts + Na-O3-contacts + Na-O6-contacts)$
while the three--body contact probability between \ce{Na+} ions, $O3$ and $O6$
is calculated as $P(Na-O3-O6)=Na-O3-O6-contacts/(Na-O3-contacts + Na-O6-contacts)$. }
\begin{ruledtabular}
\begin{tabular}{lccc}
T(K) & O2 & O3 & O6 \\
\hline
$298$ & $0.32$ & $\mathbf{0.37}$ & $0.30$ \\
$402$ & $\mathbf{0.34}$ & $0.33$ & $0.32$ \\
$502$ & $0.32$ & $0.32$ & $\mathbf{0.35}$ \\
\hline
T(K) & O2-O3 & O2-O6 & O3-O6 \\
\hline
$298$ & $\mathbf{0.22}$ & $0.02$ & $0.10$ \\
$402$ & $\mathbf{0.17}$ & $0.02$ & $0.07$ \\
$502$ & $\mathbf{0.15}$ & $0.02$ & $0.07$ \\
\end{tabular}
\end{ruledtabular}
\end{table}

\newpage

\begin{table}[h]
\caption{\label{tab:1b}
Rotational state occupancy for the hydroxymethyl group in the fibril
crystalline core and on the surface cellulose chains in cellulose I$_{\beta}$
fibrils in pure water at $298$ K (Fib-298) \cite{ShiGio2011}, in NaCl aqueous
solution at $298$ K (Fib-298-NaCl) and at $400$ K (Fib-400-NaCl).}
\begin{ruledtabular}
\begin{tabular}{lccc}
Fib-298\cite{ShiGio2011} & $tg$ & $gt$ & $gg$ \\
\hline
crystalline core & $92.8\%$ & $4.6$ & $2.6$ \\
surface chains & $25.8\%$ & $28.1$ & $46.1$ \\
\hline
Fib-298-NaCl & $tg$ & $gt$ & $gg$ \\
\hline
crystalline core & $59.9\%$ & $25.2$ & $14.9$ \\
surface chains & $15.7\%$ & $37.1$ & $47.2$ \\
\hline
Fib-400 & $tg$ & $gt$ & $gg$ \\
\hline
crystalline core & $54.5$ & $14.8$ & $30.7$ \\
surface chains & $18.1$ & $28.4$ & $53.5$ \\
\hline
Fib-400-NaCl & $tg$ & $gt$ & $gg$ \\
\hline
crystalline core & $9.3\%$ & $43.7$ & $47.0$ \\
surface chains & $11.4\%$ & $35.1$ & $53.5$ \\
\end{tabular}
\end{ruledtabular}
\end{table}

\newpage

\begin{table}[h]
\caption{\label{tab:2}
Cellulose I$_{\beta}$ fibril in NaCl aqueous solution.
Two and three-body relative contact probabilities between \ce{Na+} and cellulose
oxygens $Ox$ where $x=2,3,6$. For example, the two--body relative contact
probability between \ce{Na+} ions and $O6$ oxygens is calculated as
$P(Na-O6)=Na-O6-contacts/(Na-O2-contacts + Na-O3-contacts + Na-O6-contacts)$
while the three--body contact probability between \ce{Na+} ions, $O3$ and $O6$
is calculated as $P(Na-O3-O6)=Na-O3-O6-contacts/(Na-O3-contacts +
Na-O6-contacts)$.}
\begin{ruledtabular}
\begin{tabular}{lccc}
T(K) & O2 & O3 & O6 \\
\hline
$298$ & $0.24$ & $0.33$ & $\mathbf{0.41}$ \\
$400$ & $0.18$ & $0.35$ & $\mathbf{0.46}$ \\
\hline
T(K) & O2-O3 & O2-O6 & O3-O6 \\
\hline
$298$ & $0.05$ & $0.06$ & $0.06$ \\
$400$ & $0.08$ & $0.06$ & $0.08$ \\
\end{tabular}
\end{ruledtabular}
\end{table}

\newpage

\begin{table}[h]
\caption{\label{tab:3}
Relative contribution to the cellulose--water hydrogen bond network at $298$ K of
oxygens O2, O3, and O6. Comparison between fibril with NaCl (Fib-NaCl) and
fibril without NaCl (Fib).\cite{ShiGio2011}}
\begin{ruledtabular}
\begin{tabular}{lccc}
System & O2 & O3 & O6 \\
\hline
Fib-NaCl & $0.29$ & $\mathbf{0.32}$ & $0.27$ \\
Fib & $0.31$ & $0.24$ & $\mathbf{0.35}$ \\
\end{tabular}
\end{ruledtabular}
\end{table}

\newpage

\newpage

\end{document}


\maketitle

\begin{figure}[t]
\centering
\includegraphics[scale=0.50]{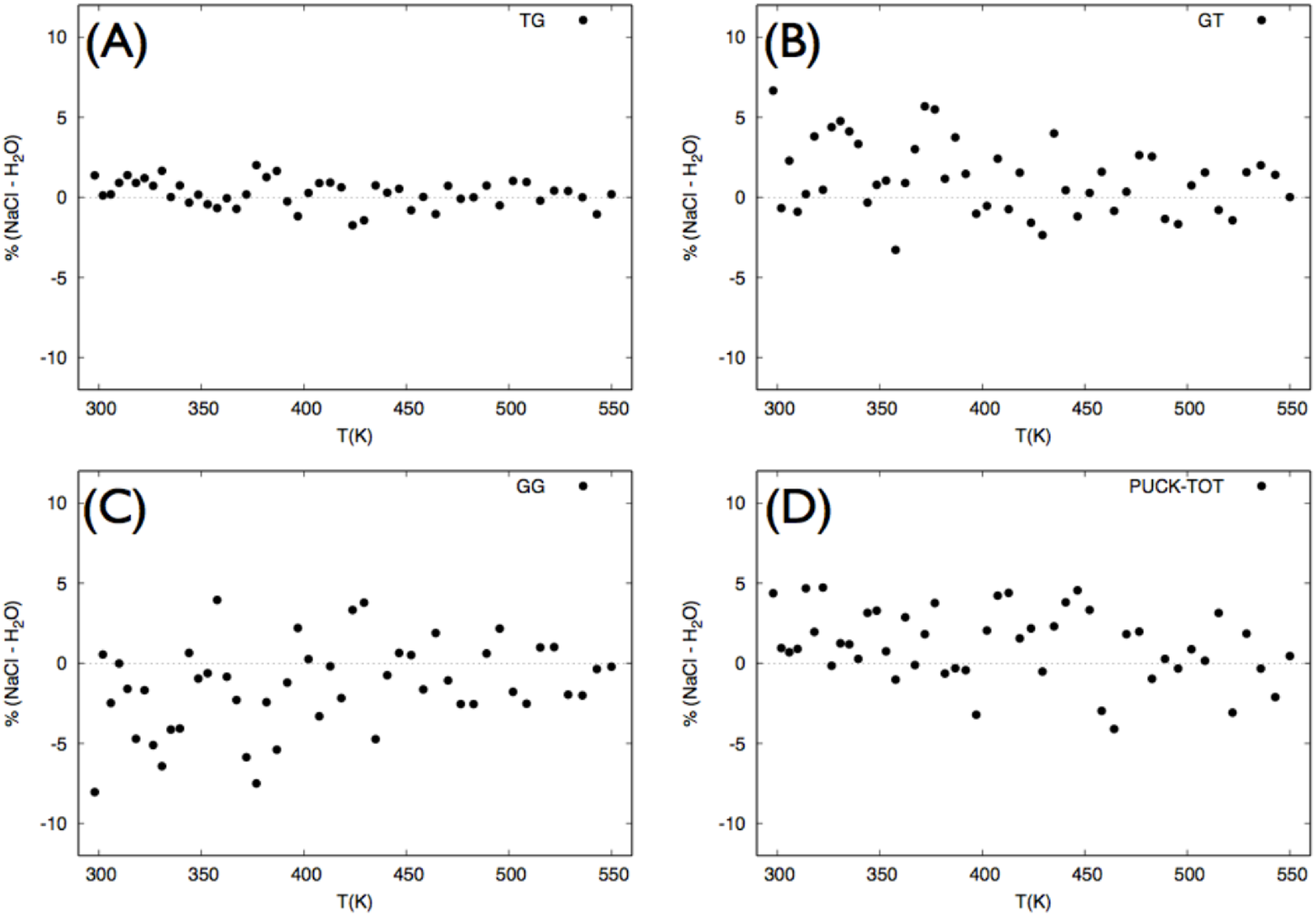}
\caption{(A), (B) and (C): residuals between cellotretrose $tg$, $gt$ and $gg$
data, respectively, calculated as the difference between the percentages
collected in NaCl aqueous solution (\ref{fig:1}(A)) and the analogous data
collected from the simulations in pure water. These data show that when NaCl is
present, the statistical weight of the $gg$ state partially diminishes mostly
at the expenses of the $gt$ state that becomes moderately more favorable than in
pure water. (D): analogous residuals for the statistical weight of the non
$^4C1$ conformations show a small increment when NaCl is present.}
\label{fig:1}
\end{figure}

\newpage
\begin{figure}[t]
\centering
\includegraphics[scale=0.50]{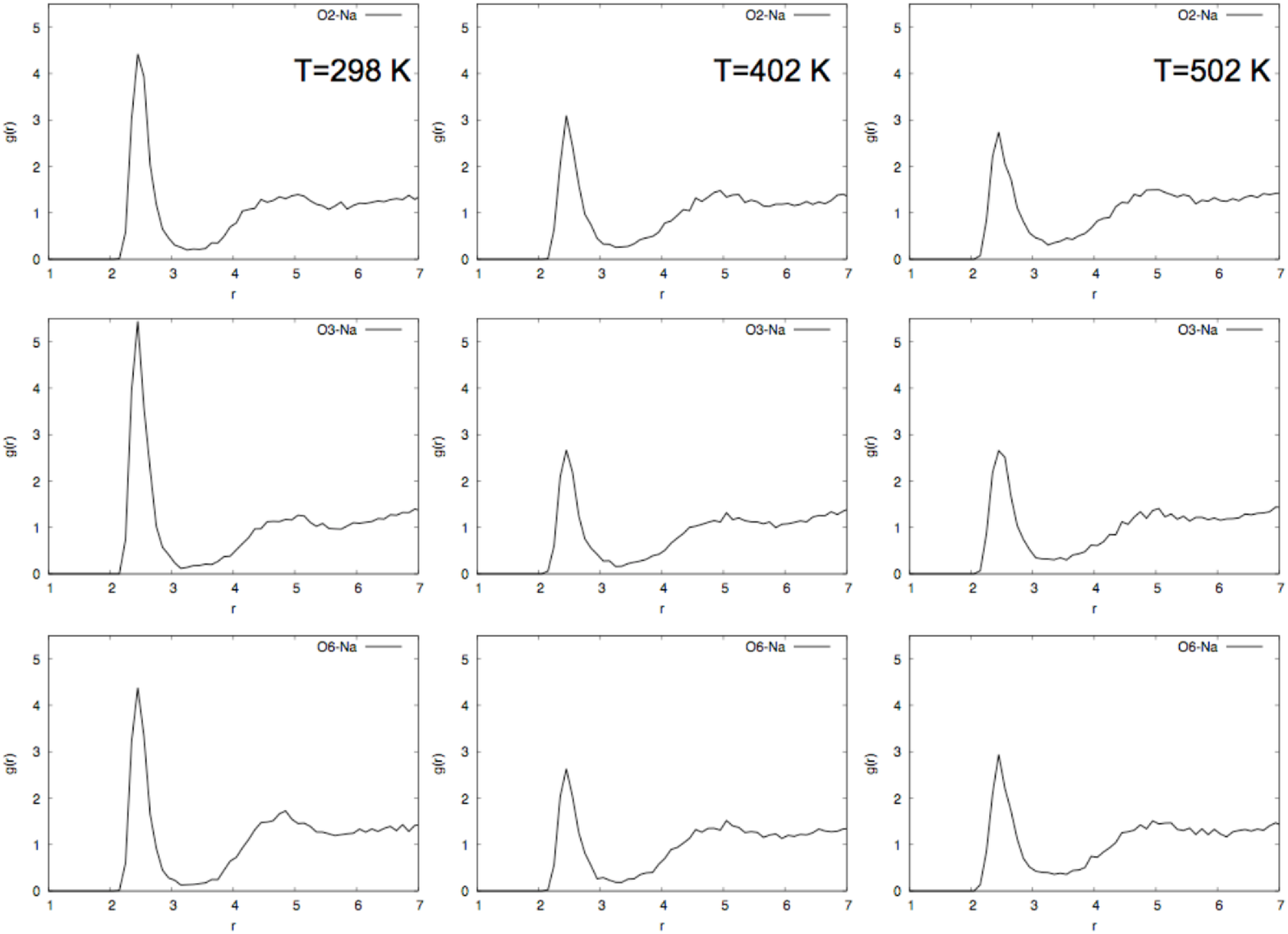}
\caption{Radial distribution functions for \ce{Na+} and $O2$, $O3$, and $O6$
(top, middle, bottom row, respectively) at 298 K, 402 K and 502 K (left, center,
right column, respectively).}
\label{fig:4}
\end{figure}

\newpage
\begin{figure}[t]
\centering
\includegraphics[scale=0.55]{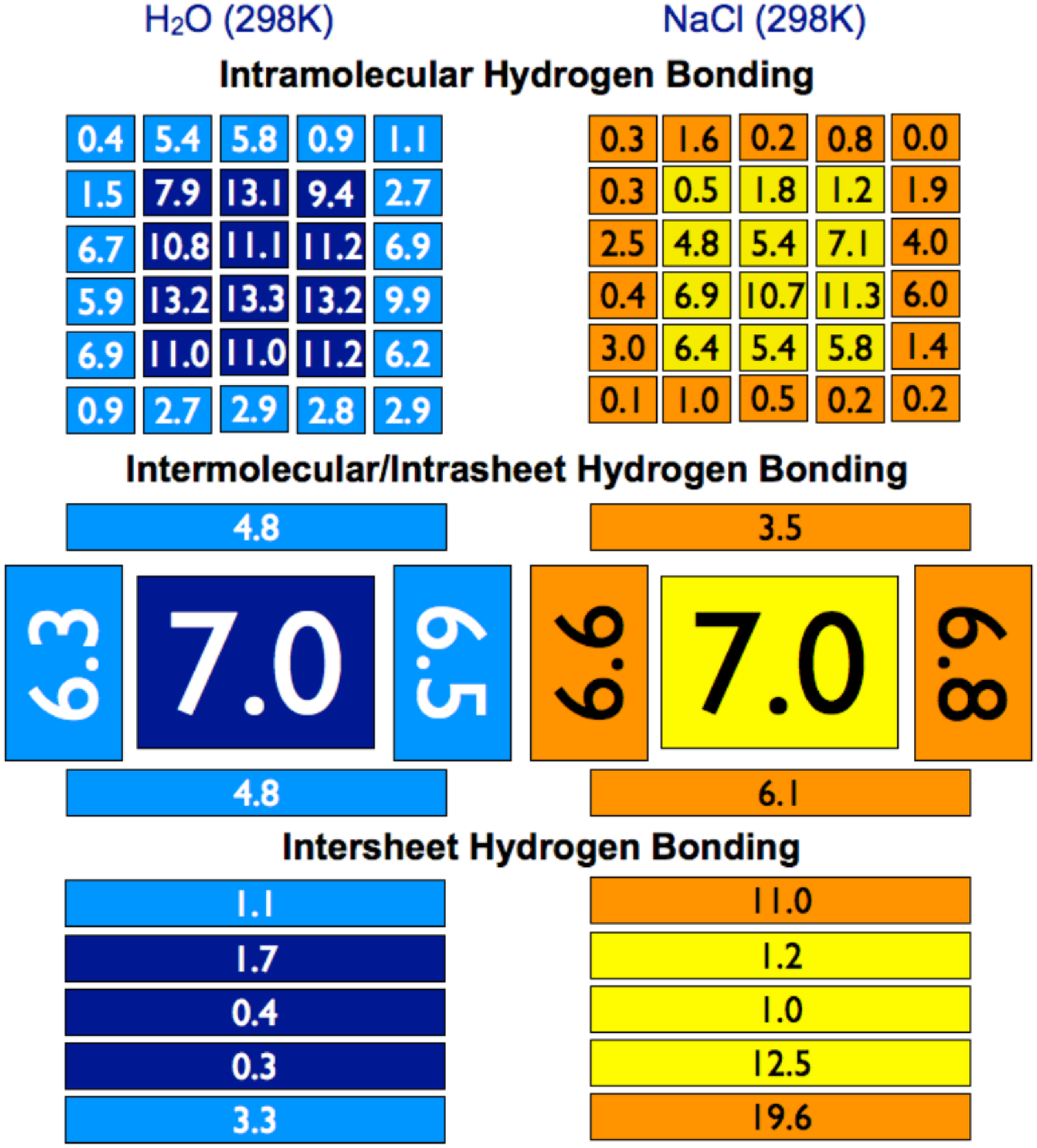}
\caption{Hydrogen bond network in cellulose I$_{\beta}$ in pure water (left)
and in NaCl solution (right) at $298$ K. The two $6x5$ grids on the top row
show the average number of intrachain hydrogen bonds per chain. The central row
shows the average number of ``interchain/intrasheet'' hydrogen bonds (between
glucan chains within the same fibril's horizontal layer), while the bottom row
shows the average number of ``interchain/intersheet'' hydrogen bonds (between
neighboring layers within the cellulose fibril).}
\label{fig:2}
\end{figure}

\newpage
\begin{figure}[t]
\centering
\includegraphics[scale=0.55]{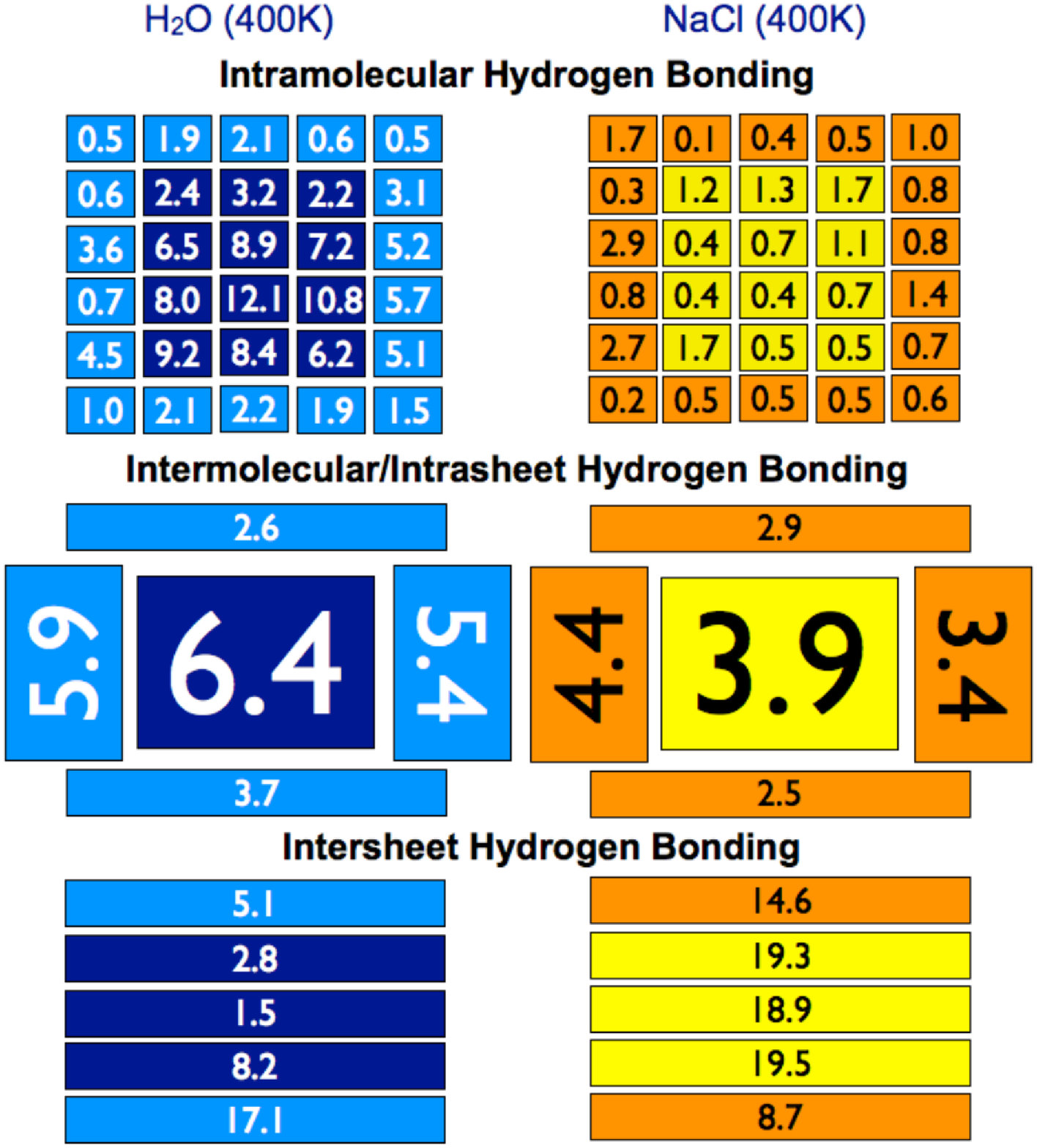}
\caption{Hydrogen bond network in cellulose I$_{\beta}$ in pure water (left) and in NaCl solution (right) at $400$ K.
The two $6x5$ grids on the top row show the average number of intrachain hydrogen bonds per chain. 
The central row shows the average number of ``interchain/intrasheet'' hydrogen bonds (between glucan chains within the same fibril's horizontal layer), 
while the bottom row shows the average number of ``interchain/intersheet'' hydrogen bonds (between neighboring layers within the cellulose fibril). }
\label{fig:3}
\end{figure}